\documentclass{aa}
\usepackage{graphicx}

\begin{document}


\title{Orbital variability of the PSR J2051$-$0827 Binary System}

\author{
        O. Doroshenko \inst{1},
        O. L\"ohmer \inst{1},
        M. Kramer \inst{2},
        A. Jessner \inst{1},
        R. Wielebinski \inst{1},\\
        A. G. Lyne \inst{2},
        Ch. Lange \inst{1}
        }

\institute{
           Max-Planck-Institut f\"ur Radioastronomie,
           Auf dem H\"ugel 69, D-53121 Bonn
           Germany
    \and
           University of Manchester, Jodrell Bank Observatory,
           Macclesfield, Cheshire SK11 9DL, UK
           }

\offprints{O.~L\"ohmer}
\mail{loehmer@mpifr-bonn.mpg.de}
\date{Received date; accepted date}

\titlerunning{Orbital variability of the PSR J2051$-$0827}
\authorrunning{O.Doroshenko {\it et al.}}

\markboth{ O.Doroshenko {\it et al.}:
 Orbital variability of the PSR J2051$-$0827}
 { O.Doroshenko {\it et al.}:
 Orbital variability of the PSR J2051$-$0827}

\abstract{
We have carried out high-precision timing measurements of the binary
millisecond pulsar PSR J2051$-$0827 with the Effelsberg 100-m radio telescope
of the Max-Planck-Institut f\"ur Radioastronomie and with the
Lovell 76-m radio telescope at Jodrell Bank. The 6.5-yrs radio timing
measurements have revealed a significant secular variation of the
projected semi-major axis of the pulsar at a rate of $\dot x\equiv
d(a_{\rm 1} \sin i)/dt = (-0.23\pm 0.03)\times 10^{-12}$, which is probably
caused by the Newtonian spin-orbit coupling in this binary system
leading to a precession of the orbital plane. The required
misalignment of the spin and orbital angular momenta of the companion
are evidence for an asymmetric supernova explosion.
We have also confirmed that the orbital period is currently decreasing at
a rate of $\dot P_{\rm b}=(-15.5 \pm 0.8)\times 10^{-12}$~s~s$^{-1}$
and have measured
second and third orbital period derivatives $d^2P_{\rm b}/dt^2=(+2.1 \pm
0.3)\times 10^{-20}\: {\rm s^{-1}}$ and $d^3P_{\rm b}/dt^3 =(3.6 \pm
0.6)\times 10^{-28}\: {\rm s^{-2}}$, which indicate a quasi-cyclic
orbital period variation similar to those found in another eclipsing
pulsar system, PSR B1957+20.
The observed variation of the
orbital parameters constrains the maximal value of the companion
radius to $R_{\rm c\: max} \sim 0.06\: R_{\odot}$
and implies that the companion is
underfilling its Roche lobe by 50 \%.
The derived variation in the quadrupole moment of the companion
is probably caused by tidal dissipation similar to the mechanism
proposed for PSR~B1957+20.
We conclude that the companion is at
least partially non-degenerate, convective and magnetically active.
}

\maketitle

\section{Introduction}

The eclipsing binary millisecond pulsar PSR~J2051$-$0827 was discovered with
the Parkes 64-m radio telescope in Australia in a 0.4~GHz survey of
the southern sky (Stappers {\it et al.} 1996a). This system has one of
the shortest known orbital periods, $P_{\rm b} \simeq 2.4$ hrs, and is moving
in an almost circular, compact orbit, with a separation between the
pulsar and its companion of only 1.03 $R_{\odot}$. Observations of the
pulsar with the 64-m Parkes and the 76-m Lovell radio telescope at
Jodrell Bank at frequencies between 0.4~GHz and 2 GHz showed that
at lower frequencies near 0.6~GHz the pulsar is eclipsed by the
atmosphere of the companion during approximately 10\% of the
orbital period, whereas at higher frequencies near 1.4 GHz there are
almost no visible eclipses in this system (Stappers {\it et al.}
1996a). High precision timing observations of PSR J2051$-$0827 revealed
that the orbital period of the system is decreasing at a rate of
$\dot P_{\rm b} \sim -10^{-11}$~s~s$^{-1}$ indicating a decay time of the
system of only 25~Myr (Stappers {\it et al.} 1998). Early optical
observations of the field of PSR J2051$-$0827 revealed that the
amplitude of the companion's light curve is about 1.2 mag and the
companion is probably rotating synchronously around the pulsar
so that one side
is being heated by the impinging pulsar flux (Stappers {\it et al.} 1996b).
The fit of photometry data to a model of a gravitationally distorted,
low-mass secondary star that is irradiated by the impinging pulsar wind,
has shown that the inclination of the system is greater than 30{\degr}
and the maximum companion mass is $0.055\: M_{\odot}$ (Stappers
{\it et al.} 1999). Recent observations of the pulsar's companion using
the Hubble Space Telescope (HST) Wide Field Planetary Camera have
allowed to detect its ``dark'' side by Stappers {\it et al.} (2001b).
Surprisingly, they detected a slight asymmetry in the
companion's light curve
indicating that a simple synchronously rotating companion
is no longer a complete model.
Fitting the same model as Stappers {\it et al.} (1999)
indicated that more than 30 \% of the pulsar
spin-down energy is converted into optical emission and that the system
is moderately inclined, i.e.~$i\sim 40^{\circ}$.

Long-term timing observations of eclipsing (and hence often interacting)
binary pulsars
can reveal additional variations in orbital elements, which are
important for understanding the evolution models of these
systems and of pulsars in general. The orbital period
derivative may change with time in a quasi-periodic way similar to
PSR B1957+20 (Arzoumanian {\it et al.} 1994) or may remain
unchanged over longer time intervals like for
PSR B1744-24A (Nice {\it et al.} 2000).

We have carried out high precision timing observations of the binary
system PSR J2051$-$0857 during the last 6.5 years. In addition to
improving previously determined timing parameters, we detect, for the
first time, time derivatives of the orbital period of second and third
order and the time derivative of the semi-major axis.
We use these discovered orbital variations to constrain 
the size of the pulsar companion. We discuss
possible additional relativistic and non-relativistic effects that 
could be the cause for the apparent variations of
the orbit.

\section{Observations}


PSR~J2051$-$0827 was observed with both the 100-m radiotelescope of
the Max-Planck-Institut f\"ur Radioastronomie in Effelsberg and the 76-m
Lovell radiotelescope at Jodrell Bank.
Observations have been carried out since 1996 using the
Effelsberg 100-m radio telescope
soon after the discovery of the pulsar
with the 64-m Parkes radio telescope in Australia.
Timing data have been acquired
approximately once per month, with a few larger gaps due to
unavailability of telescope time. We have collected about 350 individual
times-of-arrival (TOAs) at center frequencies of 0.86, 1.4, 1.7 and
2.7~GHz. Most of the observations were performed at
frequencies near 1.4~GHz.

During a period in August-October 1996, the data were obtained using the
Effelsberg Pulsar Observation System (EPOS). Here two channels of circular
polarization are processed in a polarimeter and a 4 x 60 x 666~KHz
filter bank combined with an incoherent hardware dedisperser
(Kramer {\it et al.} 1997, 1998). Only the two orthogonal
total power signals of 40~MHz bandwidth were used for
later analysis.
The input signal from the radio telescope is synchronously
accumulated into 1024 pulse bins with the apparent pulsar spin
period. The time stamp of each integration is synchronised to time signals
from a hydrogen maser clock calibrated with the signals from the
Global Positioning System (GPS). The dedispersed pulsar
profiles were integrated for 15 s and later transformed to 8 min -
integrations for further template matching in the time domain, using the
cross-correlation of the integrated profile with a high signal-to-noise
template.

Since October 1996, the data were collected using the Effelsberg -
Berkeley Pulsar Processor (EBPP), a coherent dispersion removal
processor, installed at Effelsberg (Backer {\it et al.}
1997). The EBPP provides 32 channels for both polarizations with a total
bandwidth of up to 112 MHz depending on the observing frequency and
dispersion measure.  For PSR J2051$-$0827
bandwidths of 28 MHz, 56 MHz and 112 MHz were
available at 0.9 GHz, 1.4 GHz and 2.7 GHz respectively. The outputs
of each channel are fed into dedisperser boards for coherent on-line
dedispersion.  All 64 output signals are integrated synchronously with
the apparent pulsar period.  The integrated profiles (5 min
integration time) were matched in the frequency domain (Taylor 1991) with
a synthetic template. The template itself consists of three Gaussian
components (Fig.~1 {\it right}) that were fitted to a high signal-to-noise
pulsar profile (Fig.~1 {\it left}) using the method developed by
Kramer {\it et al.} (1994, 1998).  The time stamp of each integrated
profile was provided by a hydrogen maser clock calibrated with signals from
the GPS to the Universal Coordinated Time UTC(NIST).

\begin{figure}
\centering
\includegraphics[bb=20 73 706 454,height=4.5cm]{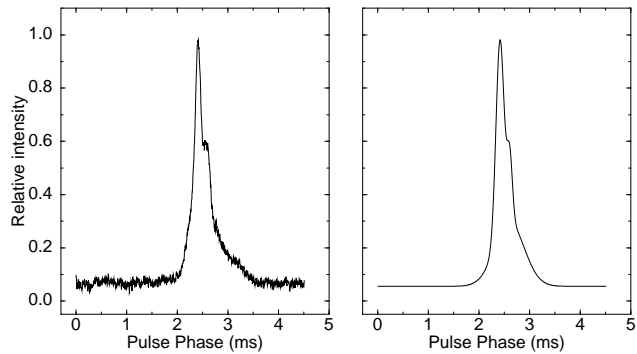}
\vspace{1.0cm}
\caption[]
{
Integrated profile of PSR J2051$-$0827, obtained with the EBPP at 1.4~GHz
{(\it left)} and the profile template, constructed from fitting of
composition of three gaussians to the integrated profile {(\it
right)}, which is used for template matching.
}
\end{figure}

In order to get a better timing solution for the orbital parameters of the
system, the sets of TOAs in several observing sessions were acquired
during one complete pulsar revolution around its orbit, equal to $\sim
2.4$ hrs.  Typical TOA uncertainties in observations are $\sim$ 8
$\mu$s at 1.4 and 1.7 GHz, $\sim 10$ $\mu$s at 2.7 GHz and $\sim$ 25
$\mu s$ at 0.9~GHz.

The Jodrell Bank data were collected since August 1994 using
cryogenic receivers at 0.3, 0.4, 0.6 and 1.4
GHz. The observations were made using both circular polarization
directions utilising a $2\times 64\times 125$~KHz filter bank.
The added
output signals from both polarizations were filtered and digitised for
further on-line hardware dedispersion and synchronous folding with
the pulsar's apparent period. Integrated profiles with integration times
of 3 minutes were matched by standard pulse template to obtain
topocentric TOAs (Bell {\it et.al} 1997). Typical TOA uncertainties
in these observations are $\sim 10$ $\mu$s at 0.4 and 1.4 GHz and
$\sim 16$ $\mu$s at 0.3 and 0.6 GHz.

\section{Data reduction and Timing solution}
 
The obtained TOAs corrected to UTC(NIST) were fitted to a spin-down
model of the pulsar rotation in the binary system using both the TIMAPR
\footnote{\small
  http://www.mpifr-bonn.mpg.de/div/pulsar/former/olegd/soft.html}
(Doroshenko \& Kopeikin 1995) and the TEMPO
\footnote{\small
    http://pulsar.princeton.edu/tempo}
software package, 
which both use the DE200
ephemeris of the Jet Propulsion Laboratory (Standish 1990) and various
relativistic timing models. Since the pulsar is moving in an almost
circular orbit, we have applied the Blandford and Teukolsky
phenomenological timing model (Blandford and Teukolsky 1976).
We have also fitted the model of Laplace-Lagrange parameters for
systems with small eccentricity (Lange {\it et al.}  2001) using
TEMPO. Comparing results obtained by TIMAPR and TEMPO are in
excellent agreement within
their uncertainties at a 1$\sigma$-level.


\begin{table}
\caption
    {Spin, astrometric, Keplerian and post-Keplerian
     parameters of binary pulsar system PSR~J2051$-$0827.
    }
\begin{tabular}{ll}
\hline
\hline \\
Parameter   &  Value   \\
\hline \\
\multicolumn{2}{c}{\it Astrometric, spin and medium parameters}\\
Right ascension $\alpha \: (J2000)$   & $20^{h} 51^{m} 07^{s}\! .514\: 5(1)$\\
Declination $\delta \: (J2000)$       & $\!\!\!\!\! -08^{\circ} 27^{'} 37^{''}\! .795(5)$\\
Proper motion in R.A.  $\mu_{\alpha} \: {\rm (mas \: yr^{-1})}$  & $  5.3(1.0)^{~a}$ \\
Proper motion in Decl. $\mu_{\delta} \: {\rm (mas \: yr^{-1})}$  & $  0.3(3.0)^{~a}$ \\
Spin period  $P$ (ms)                                   & $4.508\: 641\: 744\: 971\: 6(2)$\\
Period derivative $\dot P$ ($\rm s\ s^{-1}$)          & $1.2737(5)\times 10^{-20}$\\
Spin frequency $\nu$ (${\rm s^{-1}}$)                   & $221.796\: 287\: 344\: 250(8)$\\
Spin frequency derivative $\dot \nu$   (${\rm s^{-2}}$) & $\!\!\!\!\! -6.266(2)\times 10^{-16}$\\
Dispersion measure DM (${\rm pc\: cm^{-3}}$)            & $20.7449(4)$  \\
Derivative of DM, $\dot{\rm DM}$ (${\rm pc\: cm^{-3}\: yr^{-1}}$) & $ 0.001\: 1(3)$\\
Epoch (MJD)                                             & $51\: 000.0$ \\
\\
\multicolumn{2}{c}{\it Parameters of binary orbit}\\
Orbital period  $P_{\rm b} $ (d)                                & $0.099\: 110\: 250\: 6(2)$\\
Projected semi-major axis $x$ (lt-s)                      & $0.\: 045\: 052(2)$    \\
Eccentricity $e$                                          & $ <  0.000\: 096$      \\
Epoch of periastron $T_{\rm 0}$, (MJD)                          & $50\: 999.983\: 601\: 7(9)$\\
Longitude of periastron, $\omega_{\rm 0}$                       & $  0^{\circ}\! .0^{~b}$    \\  
Orbital period derivative $\dot P_{\rm b}$ ($\rm s\ s^{-1}$)    & $\!\!\!\!\! -15.5(8) \times 10^{-12}$ \\
Second derivative of $P_{\rm b}$, $\ddot P_{\rm b}$ (${\rm s^{-1}}$)  & $2.1(3)\times 10^{-20}$\\
Third derivative of $P_{\rm b}$, $d^3P_{\rm b}/dt^3$ (${\rm s^{-2}}$) & $3.6(6)\times 10^{-28}$\\
Derivative of semi-major axis $\dot x$ ($\rm s\ s^{-1}$)     & $\!\!\!\!\! -23(3)\times 10^{-14} $\\
\\
\multicolumn{2}{c}{\it Upper limits to parameters$^{~b}$}\\
Second derivative of $x$, $\ddot x$, (${\rm s^{-1}}$)                 & $ 7(15) \times 10^{-22}$ \\
Derivative of eccentricity $|\dot e|$ (${\rm s^{-1}}$)                & $ 1\times 10^{-12}$   \\
Second Period deriv. $ \ddot P  $ (${\rm s^{-1}}$)                    & $ 5(2)\times 10^{-31}$\\
Second frequency derivative $\ddot\nu$ (${\rm s^{-3}}$)               & $ 2(1)\times 10^{-26}$\\
\\
\multicolumn{2}{c}{\it Deduced and additional parameters}\\
Distance (DM - based), (kpc)                                  & $  1.3   $  \\
Transverse velocity $V_{\rm t} ,\  {\rm (km\ s^{-1})}$            & $ 33(19) $  \\
Shklovskii effect $\dot P_{\rm t}$ ($\rm s\ s^{-1}$)              & $ 0.01 \div 0.09\times 10^{-20}$ \\
Mass function $f_{\rm m}$ ($M_{\odot}$)                           & $0.999\: 50(1)\times 10^{-5}$\\
Companion mass $m_{\rm c}$ ($M_{\odot}$)                          & $0.0273/\sin i^{~c}$\\
\\
rms TOA residual $\sigma$ (${\rm \mu s}$)    & 21.1        \\
Timing data span (MJD)                       & 49573-51908 \\
Number of TOAs                               & 584         \\
\\
\hline 
\end{tabular}

{\small $^a$ See Fig. 4 for the values and uncertainties in proper motion.}\newline
{\small $^b$ These parameters were fixed at zero in global fit.}\newline
{\small $^c$ Calculated assuming a pulsar mass $m_{\rm p}=1.4\: M_{\odot}$.}\newline
{\small Note --- Values in parenthesis are 1$\sigma$ uncertainties in the
        last quoted digit.}
\vspace{0.2cm}
\end{table}

In the fitting procedure the three TOA segments, obtained with both
the Effelsberg and Jodrell Bank data acquisition systems, were fitted for
a mutual offset. Data lying in the range of orbital phases between
0.20 and 0.35 were excluded from the fit because of the additional modulation
of the excess column density on the TOAs in the eclipsing region (Stappers
{\it et al.} 1998, 2001a). From the fit we obtained
precise estimates of the pulsar's parameters and those of the orbit.
These include
astrometrical parameters, spin parameters, and Keplerian and
post-Keplerian orbital parameters (Table 1).

\begin{figure}
\centering
\includegraphics[bb=0 250 550 736,height=5.5cm]{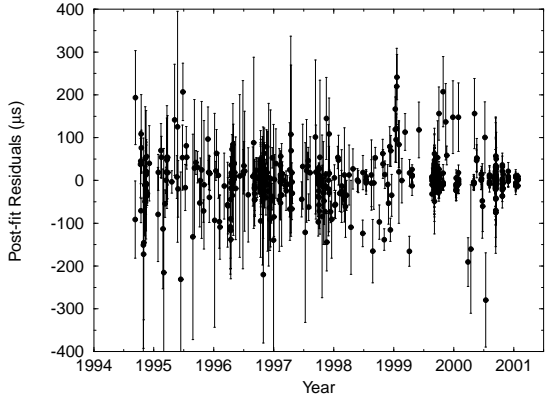}
\vspace{1.3cm}
\caption[]
{
Post-fit TOA residuals of PSR J2051$-$0827 plotted as a function of epoch,
after subtracting a global 16-parameter timing model from the data.
}
\end{figure}

Post-fit residuals are plotted as a function of observing epoch (Fig. 2)
and as a function of orbital phase (Fig. 3). The upper limits on the values 
of $\dot{\rm DM}$, $\ddot x$, $\dot e$ and $\ddot P$, presented in Table 1,
were obtained by
the individual inclusion of the corresponding parameter in the fitting
procedure. These parameters, as well as the longitude of periastron
$\omega_{\rm 0}$, were fixed at zero while fitting for the other
parameters.

\begin{figure}[h]
\centering
\includegraphics[bb=0 250 500 736,height=5.5cm]{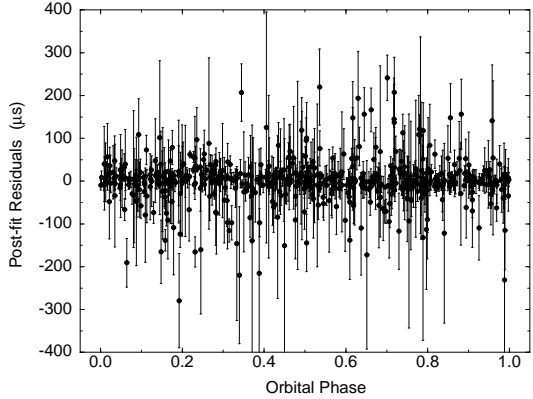}
\vspace{1.3cm}
\caption[]
{
Post-fit TOA residuals of PSR J2051$-$0827 plotted as a function of
orbital phase of the pulsar in the binary system
with $P_{\rm b}\simeq 2.4$ hrs,
after subtracting a global 16-parameter timing model from the
data. The TOAs lying in the eclipsing region with the orbital phase
(0.20-0.35) were omitted from the global fit.
}
\end{figure}


Since the proper motion in right ascension and declination
($\mu_{\alpha}, \mu_{\delta}$) has significant mutual covariances
and the TOA uncertainties are relatively large, the global
fit failed to give good estimates and uncertainties for
$\mu_{\alpha}$ and $\mu_{\delta}$. To obtain more  precise values of the
proper motion parameters, we have mapped out the region of $\chi^2$ space
near its global minimum in the ($\mu_{\alpha},\mu_{\delta}$) plane.
Contours of
$\Delta\chi^2(\mu_{\alpha},\mu_{\delta})\equiv
    \chi^2(\mu_{\alpha},\mu_{\delta}) - \chi^2_{\rm 0}$,
where $\chi^2_{\rm 0}$ is the minimum of the statistic $\chi^2$, are plotted
in Fig.~4. They show regions of 1$\sigma$ $(\Delta\chi^2 = 2.3)$,
2$\sigma$ $(\Delta\chi^2 = 6.2)$ and 3$\sigma$ $(\Delta\chi^2 =11.8)$,
which correspond to $68.3\%$, $95.4\%$ and $99.7\%$ confidence for
$\mu_{\alpha}$ and $\mu_{\delta}$.  Within 1$\sigma$ - uncertainty,
the proper motion in right ascension is $\mu_{\alpha}=5.3\pm1.0$ mas
yr$^{-1}$ and in declination is $\mu_{\delta}=0.3\pm3.0$ mas
yr$^{-1}$.  The composite proper motion of PSR~J2051$-$0827 is
$|\mu|=5.3\pm 3.0$ mas yr$^{-1}$.

\begin{figure}[h]
\centering
\includegraphics[bb=0 160 550 736,height=5.5cm]{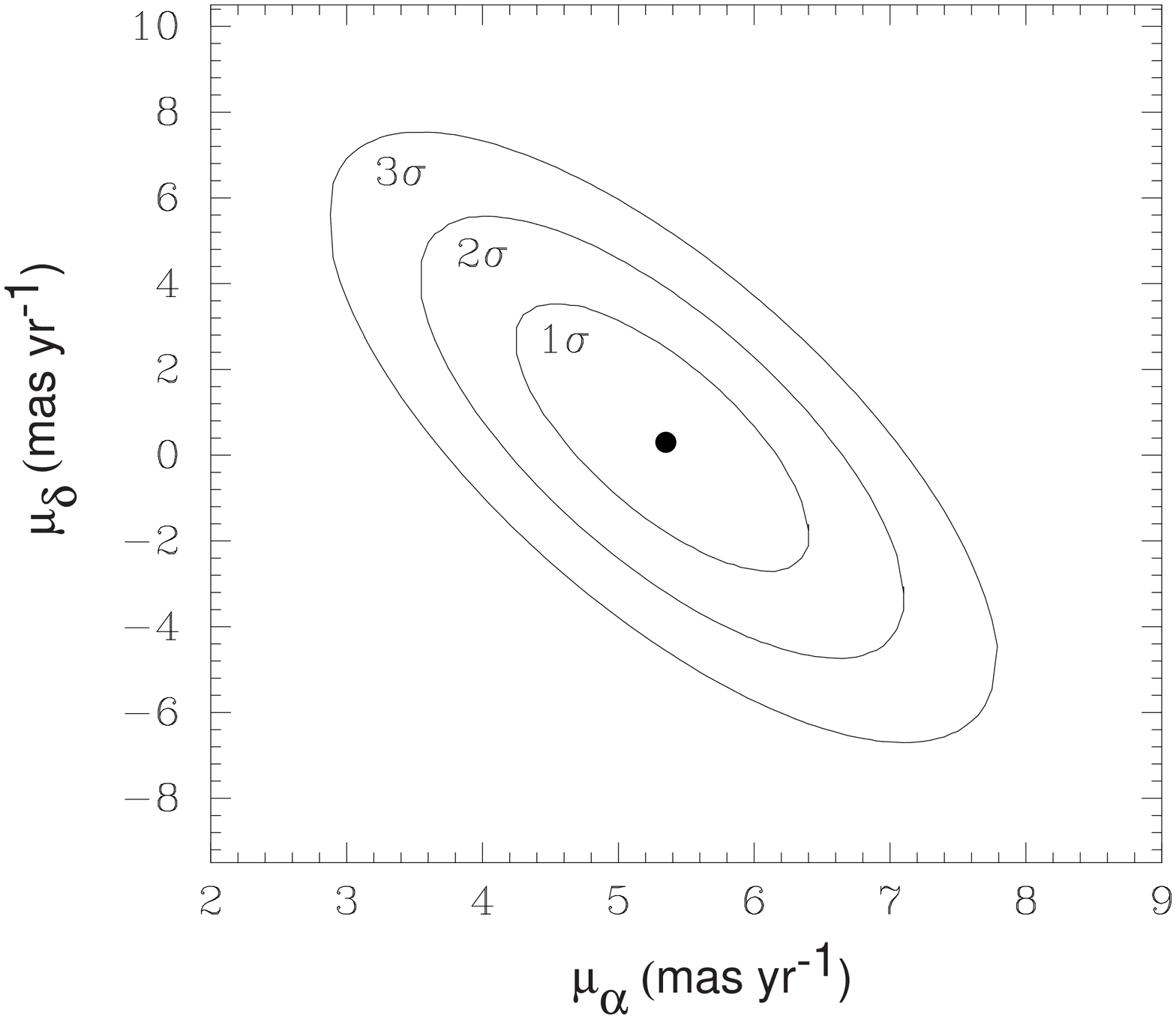}
\vspace{0.5cm}
\caption[]
{
Contours of $\chi^2$ near its global minimum in the
($\mu_{\alpha},\mu_{\delta}$) plane.
The contours enclose regions of 68.3\%,
95.4\%, and 99.7\% confidence for values of $\mu_{\alpha}$ and
$\mu_{\delta}$. The dot shows a low confidence value, corresponding
to the minimal value of $\chi^2$.
}
\end{figure}

The newly obtained celestial coordinates and binary parameters for
PSR J2051$-$0827 are
in excellent agreement to previously published values. The long
time span even allows to detect higher order time derivatives of the
orbital period 
$P_{\rm b}$ and the projected semi-major axis of the pulsar
$x\equiv a_{\rm p}\sin i$.
The measured values of $\dot P_{\rm b}$, $\ddot P_{\rm b}$
and $\dot x$ are highly significant and were detected at a level of
about $8\sigma$, reducing the post-fit TOA residuals by $\sim 25\%$.

\section{Variation of the pulsar orbit}

\subsection{Possible effects of DM variations}

Variations of the dispersion measure, DM,
which can be large in eclipsing binary systems,
can have a significant influence on the TOAs and thus
on the fitted orbital parameters and their time derivatives.
Therefore we test if the observed
variations of the orbital parameters could be induced by
DM variations.

\begin{figure}[ht]
\centering
\includegraphics[bb=0 230 550 650,height=5.5cm]{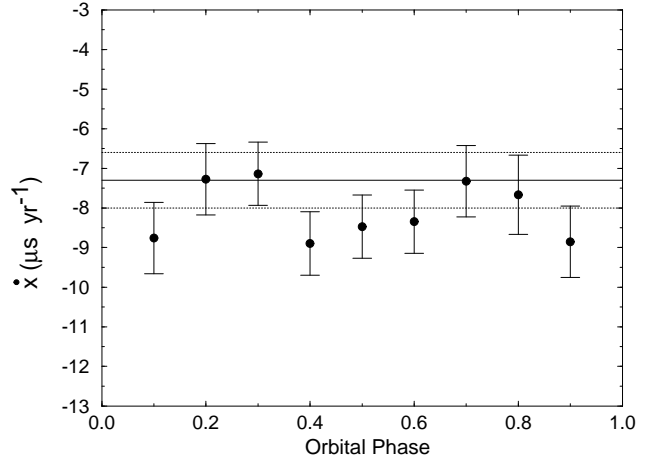}
\vspace{0.8cm}
\caption[]
{
Secular variations of the projected semi-major axis $x\equiv a_{\rm p} \sin i$
of PSR J2051$-$0827, obtained from fits of the data with
excluded corresponding orbital phases. The solid line represents
$\dot x=-7.3\: {\rm \mu s\: yr^{-1}}$ from Table 1 and the dotted
lines show the uncertainty in $\dot x$.
}
\end{figure}

The best fit solution for PSR J2051$-$0827 reveals a secular
variation of the dispersion measure at a rate
$d({\rm DM})/dt = 0.0011(1)$~pc~cm$^{-3}$~yr$^{-1}$ (Table 1). This value is
consistent with the expected square root dependence of DM variations
to the distance, caused by the motion of the pulsar through the
interstellar medium (Backer {\it et al.} 1993). Apart from this long-term
variation in DM, there might be variations over the orbit, which
are due to a non-isotropic distribution of the electron column density
$n_{\rm e}$ near the pulsar eclipsing region.
The observed value of $\dot x$
may be caused by changes of $n_{\rm e}$
near regions of upper and lower conjunction.
Such variations depend on the orbital
phase. If the assumption about the variations in $n_{\rm e}$ near
upper and lower conjunction is correct, the fit to the timing model
using the TOAs with excluded data segments in the regions of upper
and lower conjunction, should show no variation in the
projected semi-major axis $x$. Fig.~5 shows these sets of timing
solutions for $\dot x$, where we have excluded 20\% of the data
around the corresponding orbital phases
(e.g., in obtaining solution of $\dot x$ at orbital phase
0.3 we have excluded all TOAs, lying in range of phases 0.2-0.4). It
is seen that the detected secular change in $x$ does not significantly
depend onto the distribution of the electron column density along the
orbit, so that DM variations cannot explain the observed $\dot x$.

\begin{figure}[ht]
\centering
\includegraphics[bb=-10 48 547 775,height=11.5cm]{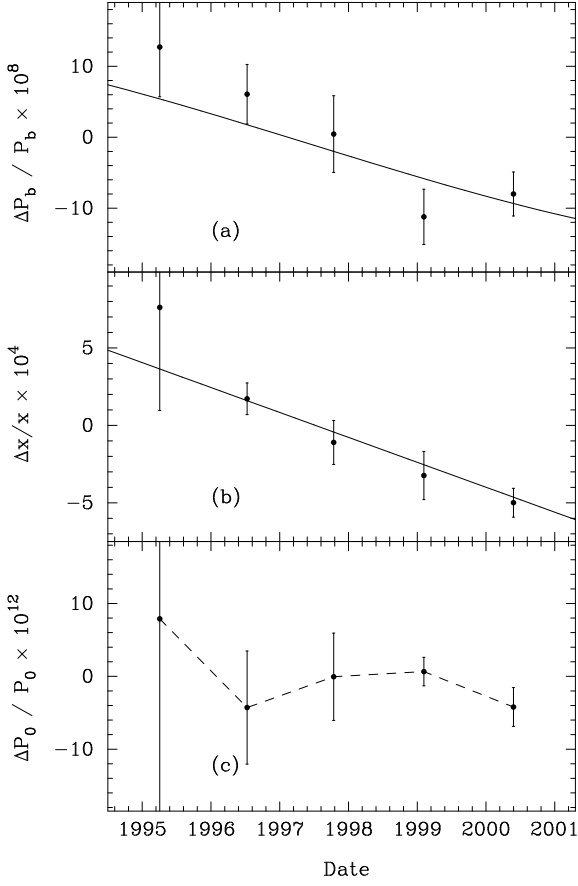}
\caption[]
{
({\it a})
Fractional changes of the orbital period in the PSR J2051$-$0827
system vs.~date, obtained from fit of the pulsar parameters
within four segments of data.
({\it b})
Fractional changes of the projected semi-major axis of PSR J2051$-$0827
vs.~date.
({\it c}) Fractional pulse period changes vs.~date.
The solid lines corresponds to the values of $\dot P_{\rm b}$,
$\ddot P_{\rm b}$
and $\dot x$ listed in Table 1.
}
\end{figure}

We have also performed a more general test of the nature of orbital
parameter variations in a way described by Arzoumanian {\it et al}
(1994). The total data set was divided into five sub-intervals, each
spanning about one year of TOAs. The individual TOA sub-sets were fitted
for the pulsar spin and Keplerian parameters with variations of orbital
parameters held fixed at zero. The time of periastron $T_{\rm 0}$ and
the orbital period $P_{\rm b}$ were transformed to
and held fixed at an epoch near the center
of each subset. The values of the pulsar spin period were transformed to
the same epoch using the spin-down model of pulsar rotation, so that
variations in period were calculated as $\Delta P_{\rm 0}\equiv \Delta P-\dot
P \Delta t$.  The resulting fractional changes in the orbital period,
the projected semi-major axis, and the spin period are shown in Figs.
6a-6c. Variations in DM
should affect measurements of the spin period $P_{\rm 0}$, orbital period
$P_{\rm b}$ and semi-major axis $x$ in the same way
as $\Delta P_{\rm b}/P_{\rm b} =
\Delta x/x=\Delta P_{\rm 0}/P_{\rm 0}$. 
As relative changes differ by a few orders of magnitude, we
conclude that the variations in DM
do not affect variations in $P_{\rm b}$, $x$ and $P$.

\subsection{Origin of the variation of the orbital period}

We have confirmed the existence of a large orbital period derivative
of the PSR~J2051$-$0827 binary system, previously
reported by Stappers {\it et
al.}  (1998). The observed value is equal to $\dot P_{\rm b} = (-15.8\pm
0.3)\times 10^{-12}\: {\rm s\: s^{-1}}$ (see Table 1) and is in good
agreement with that found by Stappers {\it et al.} (1998).

There are a number of effects that may cause the change of the
orbital period of a binary system, which can be summarised as
\begin{equation}
\label{PBDOBS}
    \left(\frac{\dot P_{\rm b}}{P_{\rm b}} \right)^{obs}\!\!\!\!\!=
    \left(\frac{\dot P_{\rm b}}{P_{\rm b}}\right)^{GW}\!\!\!\!\!\!\!+
    \left(\frac{\dot P_{\rm b}}{P_{\rm b}}\right)^{acc}\!\!\!\!\!\!+
    \left(\frac{\dot P_{\rm b}}{P_{\rm b}}\right)^{\dot m}\!\!\!\!\!+
    \left(\frac{\dot P_{\rm b}}{P_{\rm b}}\right)^{T}\!\!\!\!\!+
    \left(\frac{\dot P_{\rm b}}{P_{\rm b}}\right)^{Q}\!\!\!\! ,\;
\end{equation}
where the contributions to the observed orbital period derivative are
due to the emission of gravitational radiation ({\it GW}),
acceleration of the binary system ({\it acc}), mass loss from the system
($\dot m$), tidal dissipation of the orbit ({\it T}\,), and gravitational
quadrupole coupling ({\it }Q).

The change in orbital period due to the general relativistic orbital
decay,
\begin{eqnarray}
\label{PBDGR}
\dot P_{\rm b}^{GW} & = - (195\pi/5)
               (2\pi T_{\odot}/P_{\rm b})^{5/3}
               \left( 1+\frac{73}{24}e^2+\frac{37}{96}e^4\right) \\
\nonumber 
          &    \times  \left( 1-e^2 \right)^{-7/2}
                (m_{\rm p} m_{\rm c}) / (m_{\rm p}+m_{\rm c})^{1/3} \, \, ,
\end{eqnarray}
where $T_{\odot}\equiv GM_{\odot}/c^3=4.925490947\times 10^{-6}$ s
(Damour \& Deruelle 1986, Damour \& Taylor 1992)
and a pulsar mass of $m_{\rm p}=1.4\: M_{\odot}\: $, is expected to be in
the range $\dot P_{\rm b}^{GW} = (-6.9 \div -3.4)
\times 10^{-14}$~s~s$^{-1}$ for the allowed companion masses
$m_{\rm c} = (0.027\div0.055)\: M_{\odot}\: $.
This value is about two orders of magnitude less than the observed value of
$\dot P_{\rm b}$.

An acceleration of the binary system with respect to the Solar System
Barycentre (SSB) may be caused by the differential rotation of the Galaxy
or by a third massive body in the vicinity of the binary system. The
acceleration affects the binary period derivative $\dot P_{\rm b}$ and the
spin period derivative $\dot P$ in the same way. Even under the assumption
that the observed value of the spin period derivative is totally due to
acceleration, the contribution to the binary period derivative will be only
$(\dot P_{\rm b}/P_{\rm b})^{acc} = (\dot P/P)^{acc} = 3
\times 10^{-18}$~s$^{-1}$,
which is four orders of magnitude less than the observed value.

The mass loss from the binary system
leads to a change in the orbital period at a rate (Damour \& Taylor 1991)
\begin{equation}
\label{MLOSS}
\left(\frac{\dot P_{\rm b}}{P_{\rm b}}\right)^{\dot m}
   = \frac{8\pi^2}{m_{\rm p}+m_{\rm c}}\frac{I_{\rm a}}{c^2}\frac{\dot P_{\rm a}}{P_{\rm a}^3},
\end{equation}
where $I_{\rm a}$ denotes the moment of inertia of body $a$ ($a=p$ for the
pulsar, $a=c$ for the companion),
$P_{\rm a}$ and $\dot P_{\rm a}$ are the spin period and
period derivative. Since the companion is almost synchronously rotating
around the pulsar ($P_{\rm c}\simeq P_{\rm b}$), we can neglect its
contribution.
Using observed values of the pulsar spin period and its
time derivative, and assuming $I_{\rm p} \sim 10^{45}$ g cm$^{-2}$, the
change of the orbital period due to mass loss is $\dot
P_{\rm b}^{\dot m_{\rm p}} \sim $ 4 $\times 10^{-17}$~s~s$^{-1}$ and hence
too small to be of importance.

Tidal torques cannot be the reason for the observed
$\dot P_{\rm b}$ either, because the magnitude of the tidal
torque is orders of magnitudes too small to transfer the
necessary angular momentum (Applegate 1992, and references therein).

As all other contributions are several orders of magnitude smaller
than the observed $(\dot P_{\rm b}/ P_{\rm b})$,
we conclude that the
change of the orbital period is most likely due to gravitational quadrupole
coupling which has been proposed earlier for the eclipsing binary
system PSR B1957+20 (Applegate and Shaham 1994).
A variable
quadrupole moment which is due to a cyclic spin-up
and spin-down of the outer layers of the companion, provides a natural
explanation of the quasi-cyclic orbital variations of the PSR B1957+20
binary system found by Arzoumanian {\it et al.} (1994) and
recently confirmed by Nice {\it et al.} (2000). In fact,
our observations revealed that the orbital period derivative
of PSR~J2051$-$0827 changes with time at a rate of $\ddot P_{\rm b}=(+2.1\pm
0.3)\times 10^{-20}\: {\rm s^{-1}}$. 
This implies that the presently
decreasing orbital period will increase after $\Delta
t=-2\dot P_{\rm b}/\ddot P_{\rm b} \sim 33$ yrs and that
the system is undergoing quasi-cyclic variations like that
found for PSR~B1957+20.
The orbital period change corresponding to a variation
of the quadrupole moment of the companion, $\Delta Q$, is equal to
(Applegate and Shaham 1994)
\begin{equation}
\label{PBQ}
    \left(\frac{\Delta P_{\rm b}}{P_{\rm b}}\right)^{Q}
      = - 9 \frac{\Delta Q}{m_{\rm c} a^2}\, ,
\end{equation}
where $m_{\rm c}$ is the companion mass, and $a$ is the relative semi-major
axis of the orbit ($a=a_{\rm p}+a_{\rm c}$). The semi-major axes of the pulsar
$a_{\rm p}=x/\sin i$ and companion $a_{\rm c}=x (1.4/0.0273) /\sin i$ are
calculated for the pulsar and companion masses $m_{\rm p}=1.4\: M_{\odot}$ and
$m_{\rm c}=0.0273\: M_{\odot}$ respectively. The quadrupole moment of the
companion can be derived as (Kopal 1978)
\begin{equation}
\label{QM1}
Q = \frac{2}{9} k \frac{\Omega_{\rm c}^2 R_{\rm c}^5}{G}\, ,
\end{equation}
where $k$ is the apsidal motion constant (Claret \& Gimenez 1991),
$\Omega_{\rm c}$ is the angular velocity of the companion, $R_{\rm c}$ its
radius, and $G$ the Newtonian gravitation constant.  By
differentiating Equation ({\ref{QM1}) the variation of the quadrupole
moment with $\Omega_{\rm c}$ is
\begin{equation}
\label{DQM1}
\Delta Q = \frac{4}{9} k \frac{\Omega_{\rm c}^2 R_{\rm c}^5}{G}
                 \left( \frac{\Delta \Omega_{\rm c}}{\Omega_{\rm c}} \right) \, .
\end{equation}
From Equations (\ref{PBQ}) and (\ref{DQM1}) the variation of the
companion angular velocity corresponding to the variation in the
orbital period is derived as
\begin{equation}
\label{DQQ}
\left( \frac{\Delta \Omega_{\rm c}}{\Omega_{\rm c}} \right)
 = - \frac{1}{4} \left( \frac{\Delta P_{\rm b}}{P_{\rm b}} \right)
 m_{\rm c} a^2 \frac{G}{k R_{\rm c}^5 \Omega_{\rm c}^2} \, .
\end{equation}

We estimate that the radius of the pulsar's companion is
$R_{\rm c}\sim 0.06\: R_{\odot}$ (see Section~\ref{x-variation}).
This value is equal to half the size of
its Roche lobe $R_{\rm L} = 0.13\: R_{\odot}$ 
(Stappers {\it et al.} 1996a)
and agrees with that inferred from optical observations
of the companion (Stappers {\it et al.} 1999, 2000).

The apsidal motion constant, $k$, strongly depends on
the effective temperature of the companion. The first optical
observations indicated a temperature of $T_{\rm eff}=4000\div 4700$~K
(Stappers {\it et al.} 1996b) whereas recent observations with the HST
indicated that the backside temperature is likely to be less then 3000 K
(Stappers {\it et al.} 2000, 2001b). The corresponding value of the apsidal
motion constant is $k = 0.044\div 0.159$ (Claret 2000).
 Since the companion spin frequency is close to that
of the orbit, we can write for the general case
\begin{equation}
\label{COMPSPIN}
\Omega_{\rm c}= f\: \frac{2 \pi}{P_{\rm b}}
\end{equation}
where $f$ is close to unity. While $f = 1$ describes synchronous rotation,
the wind and the magnetic activity of the companion in the
Applegate \& Shaham (1994) model result in a
torque which slows down companion's rotation, so that $f<1$.
From the light curve observed by Stappers {\it et al.} (2001b, see
their Fig.~1), we can expect this deviation from synchronous rotation
to be small, estimating a lower limit of $f>0.9$.
For $f\approx1$, the change of the
orbital period of $\Delta P_{\rm b} \sim -1.5\times 10^{-4}\: $~s
observed after six years is caused by a variation of the
angular velocity of the companion of
\begin{equation}
\label{DQOBS}
\left( \frac{\Delta \Omega_{\rm c}}{\Omega_{\rm c}} \right)
               \sim (8\div 60) \times 10^{-4} \,
\end{equation}
for an orbital inclination $30^{\circ} < i < 90^{\circ}$ and
a range of companion masses of $m_{\rm c}=0.027\div 0.055
M_{\odot}$. For $f\approx 0.9$ we derive
$\Delta \Omega_{\rm c}/\Omega_{\rm c}\sim (1\div 74) \times 10^{-3}$.
These estimations of the angular velocity variations agree well with
$\Delta \Omega_{\rm c}/\Omega_{\rm c}\sim 10^{-3}$
for the companion of PSR B1957+20 and with
$\Delta \Omega_{\rm c}/\Omega_{\rm c} \sim (3 \div 30)\times 10^{-3}$
for Algol, RS CVn and CV systems (Applegate \& Shaham 1994).

\subsection{Origin of the variation of the projected semi-major axis
\label{x-variation}}

We have detected a significant secular variation of the pulsar
projected semi-major axis at a rate of $\dot x\equiv d(a\sin i)/dt =
(-23\pm 3)\times 10^{-14}\: \: {\rm s\: s^{-1}}$ in a global fit of
the pulsar parameters. For an independent check of this effect, the
TOA data set was split in five approximately equal segments, which were
fitted for pulsar astrometric, spin and orbital parameters with the
time derivatives of the orbital elements held fixed at zero. In these
segments the epochs of ascending node were converted to values
corresponding to the centre of each data set. The fitted values of the
projected semi-major axis for these three time intervals are found to
be in good agreement with the timing solution obtained from the
global fit, and confirm the current decrease of
the projected semi-major axis (Fig.~7).

\begin{figure}[h]
\centering
\includegraphics[bb=10 100 633 520,height=5.5cm]{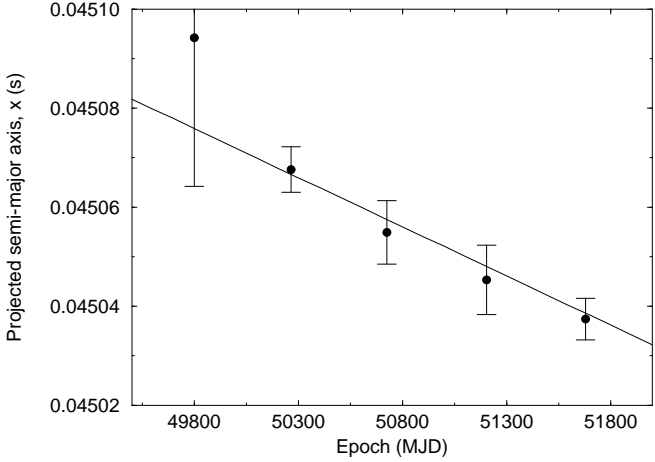}
\vspace{1.0cm}
\caption[]
{
Secular variation of the projected semi-major axis $x\equiv a_{\rm p} \sin
i$ of PSR J2051$-$0827. The dots indicate the values with their
formal errors,
obtained from fits of subsequent TOA data segments with the
time derivatives of the orbital elements held fixed at zero.
The solid line
shows the secular variation of $x$ obtained from the global fit.
}
\end{figure}

The value of $\dot x$ cannot be explained by the observed time
derivative in the orbital period $\dot P_{\rm b}$. The observed value of
$\dot P_{\rm b}$ assumes a corresponding change in the projected semi-major
axis as a consequence of Kepler's third law, at a rate
\begin{equation}
\label{XDPD}
         \frac{\dot x}{x} = \frac{2}{3} \frac{\dot P_{\rm b}}{P_{\rm b}} .
\end{equation}
For the measured values of $x$, $P_{\rm b}$ and $\dot P_{\rm b}$
one obtains
$\dot x_{\rm max}\simeq -6\times 10^{-17} {\rm s\: s^{-1}}$, which is about
4 orders of magnitude less than the observed $\dot x$.

There are a few effects which may cause a secular change of the
projected semi-major axis $x$ (Damour \& Taylor
1992, Kopeikin 1994, Wex \& Kopeikin 1999), i.e.
\begin{equation}
\label{XDOBS}
    \left(\frac{\dot x}{x} \right)^{obs}\!\!\!\!=
    \left(\frac{\dot a_{\rm p}}{a_{\rm p}}\right)^{GW}\!\!\!+
    \left(\frac{\dot x}{x}\right)^{PM}\!\!\!+
    \frac{d\varepsilon_{\rm A}}{dt}-
    \frac{\dot D}{D} +
    \left(\frac{\dot x}{x}\right)^{SO}\; ,
\end{equation}
where the contributions to the observed semi-major axis derivative are
due to the emission of gravitational radiation ({\it GW}),
proper motion of the binary system ({\it PM}),
varying aberration $(d\varepsilon_{\rm A}/dt)$,
changing Doppler shift ($-{\dot D}/D$) and spin-orbit
coupling ({\it SO}) in the binary system.

The first contribution ({\it GW}) in Equation~(\ref{XDOBS})
represents the shrinking of the pulsar orbit due to
gravitational-wave damping in the binary system given by
\begin{equation}
\label{ADGW}
\left(\frac{\dot a_{\rm p}}{a_{\rm p}}\right)^{GW}
        = \frac{2}{3}\left(\frac{\dot P_{\rm b}}{P_{\rm b}}\right)^{GW}.
\end{equation}
With $\dot P_{\rm b}^{GW}$ calculated from Equation (\ref{PBDGR}), one
expects a rate of 
$(\dot a_{\rm p}/a_{\rm p})^{GW} \sim (-5.4 \div -2.6)\times 10^{-18}$~s$^{-1}$,
which is about 6 orders of magnitude smaller than the observed
$(\dot x/ x)$ and can be neglected.

The second contribution ({\it PM}) in Equation (\ref{XDOBS}) is caused
by the proper motion of the binary system (Kopeikin 1996,
Arzoumanian {\it et al.}  1996; Bell {\it et al.} 1997) given by
\begin{equation}
\label{XDPM}
    \left(\frac{\dot x}{x}\right)^{PM}\!\!\!\!\! = 1.54\times 10^{-16}\: \cot i\:
             (-\mu_{\alpha}\sin \Omega +\mu_{\delta}\cos \Omega)\: ,
\end{equation}
where $\Omega$ is the longitude of the ascending node of the orbit,
and $\mu_{\alpha}$ and $\mu_{\delta}$ (mas yr$^{-1}$) 
are the proper motions in right
ascension and declination. In this case, the
observed change in $x$ is caused by the secular variation in the
inclination angle $i$ due to the proper motion, whereas the intrinsic
semi-major axis of the pulsar $a_{\rm p}$ remains constant. The maximal
contribution of the proper motion is
\begin{equation}
  \left(\frac{\dot x}{x}\right)^{PM}_{\rm max}
   \simeq 1.54\times 10^{-16}\: \mu\: \cot i
\end{equation}
with a composite proper motion
$\mu=(\mu_{\alpha}^2+\mu_{\delta}^2)^{1/2}= 5(3)$ mas yr$^{-1}$.
In order to estimate this quantity
we still need an
estimation of the maximal inclination angle $i$.

If the
companion is a white dwarf, we obtain a minimal radius using
the mass-radius relation of Paczy\'nski (1967) of
$R_{\rm c\:min}=0.03\:R_{\odot}$ for mass ranges of
$m_{\rm c}=0.027\div 0.055\: M_{\odot}$. 
If the companion is, instead, a non-degenerate star in the pulsar
radiation field, as suggested by Ergma {\it et al.} (1998),
the radius is even larger.
We hence obtain $R_{\rm c\:min}=0.03\:R_{\odot}$ as a safe
lower limit. Since at higher frequencies
the pulsar signal is not eclipsed by the companion that is
separated by $1.03\:R_{\odot}$, we derive
an upper limit for the inclination angle of 
$i_{\rm max}=88^{\circ}$.
This inclination angle leads to
$\dot x^{PM}_{\rm max} = 1.2(7)\times 10^{-18}$~s~s$^{-1}$. For
the likely inclination of $i\sim 40^{\circ}$
(Stappers {\it et al.} 1999)
we obtain $\dot x^{PM} = 4(2)\times 10^{-17}$~s~s$^{-1}$.
This is at least four orders of magnitude smaller than the observed
$\dot x$ and can be neglected.

The third term, $(d\varepsilon_{\rm A}/dt)$, in Equation (\ref{XDOBS}) is due
to varying aberration caused by relativistic precession of the pulsar
spin axis
\begin{equation}
\label{DEDT}
 \frac{d\varepsilon_{\rm A}}{dt} =
 - \frac{P}{P_{\rm b}}\; \frac{\Omega_{\rm p}^{geod}}{(1-e^2)^{1/2}}
   \frac{(cot \lambda \sin 2\eta + \cot i \cos \eta)}
        {\sin\lambda}\: ,
\end{equation}
where $P$ is the pulsar spin period, and ($\eta, \lambda$) are the
polar coordinates of the pulsar spin (see Damour \& Taylor 1992).
The rate of geodetic precession given by (Barker and O'Connel 1975)
\begin{equation}
\label{OMGEOD}
\Omega_{\rm p}^{geod} = \frac{G^{2/3}}{2 c^2 (1-e^2)}
                  \left( \frac{2\pi}{P_{\rm b}} \right)^{5/3}
                  \frac{m_{\rm c} (4 m_{\rm p}+3 m_{\rm c})}{(m_{\rm p}+m_{\rm c})^{4/3}}\; ,                 
\end{equation}
amounts to $\Omega_{\rm p}^{geod} \sim 1.7\times
10^{-10} = 0.3^{\circ}$ yr$^{-1}$ for a companion mass of $m_{\rm c}=0.055\:
M_{\odot}$. This leads to $(d\varepsilon_A/dt) < 8.5\times 10^{-20}$~s$^{-1}$,
which is about 6 orders of magnitude less than the observed
$(\dot x/ x)$ and can be neglected.

The fourth term in Equation (\ref{XDOBS}), ($\dot D/D$), is due to a
changing Doppler shift (D) caused by the change of the distance between the
SSB  and the binary pulsar system due to
the acceleration of the binary system
in the gravitational field of the Galaxy and due to
the proper motion on the plane of the sky (Shklovskii
1970, Damour \& Deruelle 1986). We obtain
\begin{equation}
\label{DOPD}
    -\frac{\dot D}{D} =
           \frac{1}{c} \: {\bf K}_{\rm 0} \cdot ({\bf a}_{\rm PSR} -
  {\bf a}_{\rm SSB})
          + \frac{V_{\rm T}^2}{c\: d}\; ,
\end{equation}
where ${\bf K}_{\rm 0}$ is the unit vector of the pulsar along the line of
sight, ${\bf a}_{\rm PSR}$ and ${\bf a}_{\rm SSB}$ are the 
Galactic accelerations
at the location of the PSR~J2051$-$0827 binary system and the SSB,
$V_{\rm T}$ the transverse
velocity of the pulsar, and $d$ the distance between the pulsar and
the SSB. Using an estimation of
$a_{\rm SSB} \sim 2.1 \times 10^{-7} {\rm m\: s^{-2}}$
(Carlberg \& Innanen 1987) we derive for the 
first term in Equation (\ref{DOPD}) an upper limit of
$7\times 10^{-16}\: {\rm s^{-1}}$. The
transverse velocity of the pulsar, $V_{\rm T} = 33\, {\rm km\, s^{-1}}$, and
a distance to the pulsar of $d = 1.3\: $ kpc yield a value of 
$V_{\rm T}^2/c\: d \sim 10^{-19}\: {\rm s^{-1}}$. Thus,
the total contribution from a varying Doppler shift to the observed
$(\dot x/ x)$ is less than $0.3 \%$ and can be neglected.

As all other contributions are several orders of magnitude smaller
than the observed $(\dot x/ x)$, the variation of $x$ is
most likely caused by classical spin-orbit coupling (SO) in the
binary system (Equation \ref{XDOBS}). Spin-orbit coupling leads to a
variation in the inclination angle, ($di/dt$), while the
semi-major axis $a_{\rm p}$ remains constant.
In this case the observed value $(di/dt)$ is
\begin{equation}
\label{IDSO}
    \frac{di}{dt} = \left(\frac{\dot x}{x}\right)^{SO} \tan i \: .
\end{equation}
If we assume that the observed $\dot x$
is totally due to spin-orbit coupling, we obtain
$(di/dt)= -0.008^{\circ}$ yr$^{-1}$ for an orbital inclination of
$i=40^{\circ}$ and $(di/dt)= -0.26^{\circ}$ yr$^{-1}$ for
$i=88^{\circ}$.

The rotationally induced quadrupole of the companion produces a
variation of the orbital inclination angle
as (Lai {\it et al.} 1995, Wex 1998)
\begin{equation}
\label{DIDTQ}
 \frac{di}{dt} =  \frac{1}{2}\: \Omega_{\rm orb}\:
 \frac{k\: R_{\rm c}^2 \: \hat {\Omega}_s^2}{a^2 (1-e^2)^2}\:
             \sin (2\theta) \sin \Phi\: .
\end{equation}
Here $\Omega_{\rm orb}\equiv2\pi/P_{\rm b}$ is the orbital frequency
and $\hat {\Omega}_{\rm s}\equiv \Omega_{\rm c}/(Gm_{\rm c}/
R_{\rm c}^3)^{1/2}$ the dimensionless
spin of the companion, $R_{\rm c}$  its radius, $a$ the semi-major axis
of the relative orbit, k the apsidal motion constant, $\theta$
the angle between the companion spin angular momentum {\bf S} and the
orbital angular momentum {\bf L}, and $\Phi$ the orbital plane
precessional phase. Equation (\ref{DIDTQ})
is valid for the case that $|{\bf S}| \ll |{\bf L}|$, which is true
for PSR J2051$-$0827, where $|{\bf S}|\sim 10^{-3} |{\bf L}|$. 
Note that spin-orbit coupling requires the companions's spin axis
to be inclined with respect to the orbital angular momentum vector
($\theta\neq 0$).
Measurement of ($di/dt$) and the classical periastron advance
$\dot\omega$ due to spin-orbit coupling can in principle be
used to obtain constraints of the values of $\theta$
and $\Phi$ (Kaspi {\it et al.} 1996, Wex 1998).
Unfortunately, the PSR J2051$-$0827 system has a negligible
small eccentricity and all effects of secular variations in
$\omega$, along with the substantial relativistic
advance of periastron $\dot \omega^{GR} \sim 12 ^{\circ}$~yr$^{-1}$,
are fully absorbed by the redefinition of the binary period.
Therefore, they are not observable in this system (Kopeikin \&
Ozernoy 1999), so that we cannot obtain constraints from
measurements of $\dot \omega$. Nevertheless we can yield an upper limit
for ($di/dt$) and thus for the companion's radius from
equations (\ref{IDSO}), i.e.
\begin{equation}
\label{DISOMAX}
\left(\frac{di}{dt}\right)_{\rm max} =
       \left(\frac{\dot x_{\rm max}}{x_{\rm min}}\right)
       \tan i_{\rm max}\: ,
\end{equation}
where the indices max, min refer to the maximal or minimal value of
the corresponding parameter substituting observed values of
$\dot x_{\rm max}$ and $x_{\rm min}$ from Table 1.
Using the maximal inclination $i_{\rm max}=88^{\circ}$,
the corresponding upper limit $(di/dt)_{\rm max}
\sim -1.7\times 10^{-10} \, {\rm s}^{-1}$,
$k=0.044\div 0.159$, \ 
$a=1.03\: R_{\odot}$ and $e=0$, we then obtain for the case of a
synchronously rotating companion ($\Omega_{\rm c}=\Omega_{\rm orb}$)
a maximal companion radius of $R_{\rm c\: max}\sim 0.05\: R_{\odot}$.
According to optical observations,
the system is likely to be moderately inclined with
an inclination angle $i \sim 40^{\circ}$ (Stappers {\it et al.} 2000, 
2001b) which would lead to $(di/dt)_{\rm max}
\sim -4.8\times 10^{-12}$~s$^{-1}$ 
and a maximal companion radius of $R_{\rm c\: max}\sim
0.03\: R_{\odot}$. The latter value is even smaller than the {\em miminal}
radius for a pure helium white dwarf companion 
(Paczy\'nski 1967) and therefore not very likely.
If the companion is not perfectly synchronously rotating 
($f\approx 0.9$, see Equation~\ref{COMPSPIN}), we obtain
$R_{\rm c\: max}\sim 0.04\: R_{\odot}$ for $i \sim 40^{\circ}$ and
$R_{\rm c\: max}\sim 0.06\: R_{\odot}$ for $i \sim 88^{\circ}$. 
A radius of $R_{\rm c}\sim 0.06\: R_{\odot}$
is equal to half the size of
its Roche lobe $R_{\rm L} = 0.13\: R_{\odot}$ 
(Stappers {\it et al.} 1996a)
and agrees well with that inferred from optical observations
of the companion (Stappers {\it et al.} 1999, 2000).

\section{Discussion and Conclusions}

The detected quasi-cyclic orbital variations of the
PSR~J2051$-$0827 binary system are most likely caused by a variable
quadrupole moment of the companion.  For PSR~B1957+20 Applegate \&
Shaham (1994) suggested that the companion's wind due to the pulsar
irradiation and its magnetic activity result in a torque that forces
the companion slightly out of synchronous rotation. We propose that
this mechanism is the likely source of orbital variability also in the
PSR~J2051$-$0827 binary system.  In this framework the resulting tidal
dissipation of energy is the source of the magnetic activity that
causes the cyclic spin-up and spin-down of the outer layers of the
companion. This makes the companion of PSR~J2051$-$0827 the only second
identified tidally powered star. We conclude that the companion is at
least partially non-degenerate, convective and magnetically active
(cf.~Arzoumanian {\it et al.} 1994) in accordance with a model
proposed by D'Antona \& Ergma (1993).

The torque acting on the companion in the Applegate \& Shaham (1994)
model causes small deviations from a perfect co-rotation ($f<1$). This
situation may be an explanation for the slight asymmetry in the
companion's light curve that was recently observed by Stappers {\it et
al.} (2001b). We note, however, that even though a similar asymmetry
detected for the companion of PSR B1957+20 by Djorgovski \& Evans
(1988) could not be confirmed by later observations (Callanan {\it et
al.}~1989), it was first interpreted as the result of a shock caused
by the interaction of the pulsar and companion winds. A similar
situation may be present in PSR~J2051$-$0827, but the non-perfect
co-rotation could well play an important role in deciphering the
observed light curve.

The variation of the projected semi-major axis of the pulsar can be
explaind by classical spin-orbit coupling. Due to a significant
quadrupole moment of the companion the spin and orbital angular
momenta, which are not aligned, couple and precess about the fixed
total angular momentum vector. The precession of the orbit results in
the observed variation of the inclination angle whereas the periastron
advance cannot be detected due to the extremely small eccentricity of
the system. The misalignment of the present system provides evidence
that the neutron star received a kick at birth (cf.~Kaspi {\it et al.}
1996) presenting further proof for the existence of asymmetric
supernova explosions. A more detailed study will be presented
elsewhere (L\"ohmer {\it et al.}, in prep.).

Using timing information we are able to obtain a maximal
radius of the companion of $R_{\rm c\: max}\sim 0.06\: R_{\odot}$, which
is about half the size of its
Roche lobe $R_{\rm L}=0.13\: R_{\odot}$. This value is consistent
with that obtained by modelling the light curve of the
companion (Stappers {\it et al.} 2001b).

The quasi-cyclic variations of the orbital period show that there is
no secular orbital decay leading to a destruction of the binary
system, which has been discussed, based on previous timing data, by
Stappers {\it et al.} (1998).  As there is no Roche lobe overflow and
the timescale to evaporate the companion by the pulsar's relativistic
wind is $\sim\: 10^9$~yr (Stappers {\it et al.} 1998) the
PSR~J2051$-$0827 binary system is not likely to become an isolated
millisecond pulsar in the future.  

\begin{acknowledgements}
We are very grateful to all people involved in the pulsar timing project
in Effelsberg, in particular to Karl
Grypstra for maintaining the observatory clocks, and to the telescope
operators at Effelsberg. We thank Norbert Wex and Aris Karastergiou
for their assistance in observations.  O.D. gratefully acknowledges
the Alexander von Humboldt-Stiftung for supporting this work and
acknowledges the receipt of the Max-Planck-Gesellschaft contract.
\end{acknowledgements}


\end{document}